\newcommand{\Msun}{$M_\odot\,$}
\newcommand{\gtsim}{\stackrel{\textstyle>}{\raisebox{-0.7ex}{$\sim$}}\,} % ~>
\documentstyle[epsf,psfig]{aipproc}
\begin{document}

\title{The Impact of the UV Upturn on the Optical Color Evolution of E Galaxies}

\author{Sukyoung Yi}

\address{NASA/GSFC, Code 681, Greenbelt, MD 20771, USA}

\maketitle

\begin{abstract}

   The UV upturn in giant elliptical galaxies (gEs) implies significant 
corrections to the model-predicted optical colors of distant galaxies. 
However, since the origin of the UV upturn is still not clear, observers 
often use models that may not be acceptable. 
   We show that ``the HB hypothesis'' of Yi et al. (1997) that explains 
most of the empirical constrains predicts a significantly different optical 
color evolution from some popular models.
   We can test this model against others using modern data.
   Meanwhile, it seems dangerous to draw serious conclusions on cosmology
when the analysis is based on oversimplified models, such as 
$k$-correction-only models or $k$+$e$ correction models that do not explain 
the UV upturn.

\end{abstract}

% section headings given below are just a guide-- edit to your own style.
% You need not even divide the text up into sections.  Tables and figures
% can be inserted using the instructions in the AIP's ``toolkit'', 
% available from their website, and any standard LaTeX cookbook.

\section*{Introduction}

   Colors provide information on the age and metallicity of a stellar
population, while the age and metallicity of galaxies as functions of redshift
constrain cosmology.
   As modern data on gEs reach farther in space and time, one can, in 
principle, put stronger constraints on cosmological theories.
   However, one now has to deal with the rest-frame UV spectrum whose 
evolutionary behavior is still poorly understood.
   The UV upturn phenomenon \cite{bur,cw} in gEs has been shown to cause 
significant corrections to models of optical (in observer's frame) color 
evolution of galaxies \cite{chi,grv}. 
   However, the cause of the UV upturn is still being debated, and different 
hypotheses that explain the UV upturn predict significantly different UV 
spectral evolution of galaxies. 
   For example, some models suggest that a UV upturn is most likely 
caused by hot horizontal-branch (HB) stars (``the HB hypothesis'') and thus 
is sign of an old population \cite{dor,tan,ydo1}.
   Other models, with young main-sequence (MS) stars or post-asymptotic giant 
branch (PAGB) stars as the primary UV source \cite{bc,gst}, do not 
predict any strong age-dependence for the UV upturn.
   As a result, many observers match their data with model colors that have 
been built without considering the possible UV spectral evolution for gEs,
e.g., correcting only for redshift. 
   Using Yi's recent models which are based on the HB hypothesis and that 
reasonably match empirical data \cite{ydo1,ydo2}, we present new optical color 
evolution models and compare them to the models based on conventional 
assumptions. 
   Fitting modern observational data with such models will not only select the 
most plausible hypothesis as the cause of the UV upturn but also allow us to 
construct more accurate color evolution models that are crucial to 
understanding cosmology.

\section*{Impact of the UV upturn}

   The significance of the UV upturn to the (observer's frame) optical color 
evolution is obvious. The question is whether the UV upturn as seen in nearby 
gEs is always present or whether it is a sign of evolution.
   Four models are presented in Figures 1 \& 2.
   They are all based on the same population model.
   In each case, we assume an instantaneous initial star burst, an 
infall-based metallicity distribution \cite{tan}, a Salpeter initial mass 
function, $\Delta$$Y$/$\Delta$$Z$ = 2, a mass loss efficiency parameter in 
Reimers' formula $\eta \approx$ 0.7, and a Gaussian-dispersion parameter on 
the estimated mass loss $\sigma$ = 0.06 \Msun. 
   The choice of parameters is discussed in Yi et al.'s papers 
\cite{ydo1,ydo2}.
   In case of $k$+$e$, $k$, $fixed$-$UV$ models listed below, models at 
$z$ = 0 fit the overall (far-UV to IR) spectrum of NGC\,4552 
(one of the UV-strong gEs), assuming an age of $\approx$ 16 Gyr.
   The $flat-UV$ model is identical to the $k$+$e$ model except that it does
not fit the UV spectrum of NGC\,4552 because it does not show a UV upturn.
   We assume ($H_o$, $q_o$, $\Lambda$) = (50, 0.05, 0).

\begin{figure}[t]
\centerline{\epsfxsize=5.5in \epsffile{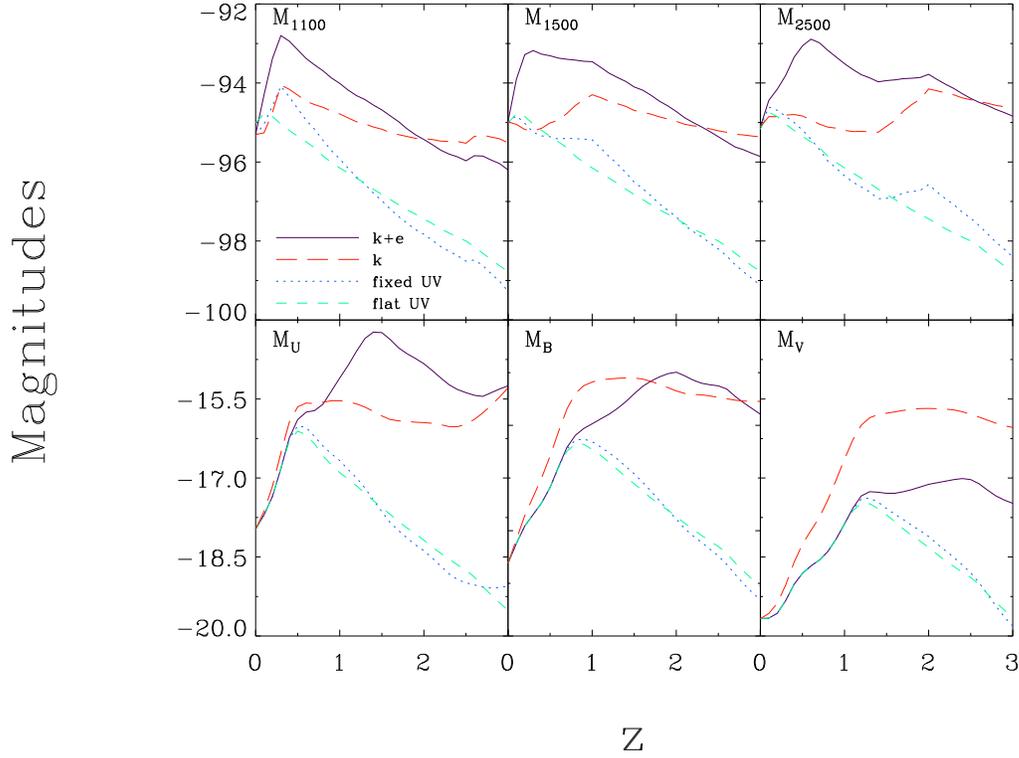}}
\vspace{15pt}
\caption
{
The luminosity evolution of gEs as a function of redshift. The $UV$ magnitudes
are defined as $M{(\lambda)}$~$=$~$-2.5$~log~$<\!f{(\lambda)}\!>$, where 
$<\!f(1100)\!>$, $<\!f(1500)\!>$, and $<\!f(2500)\!>$ are the mean flux within
the ranges 1050 -- 1200 \AA, 1250 -- 1850 \AA, and 2200 -- 2800 \AA.
The $k$ and $e$ stand for $k$- and $e$-corrected. 
The $fixed$-$UV$ and the $flat$-$UV$ models are both $k$+$e$ corrected, but 
the former assumes that a UV upturn always exists with the same relative 
strength to $f({\lambda= 2500})$ as seen in NGC\,4552 while the latter assumes 
that a UV flux is always flat (i.e., $f{(\lambda<2500)} = f{(\lambda=2500)}$).
The $Y$-axis scale is arbitrary.
}
\end{figure}
 
\begin{figure}[t]
\centerline{\epsfxsize=5.5in \epsffile{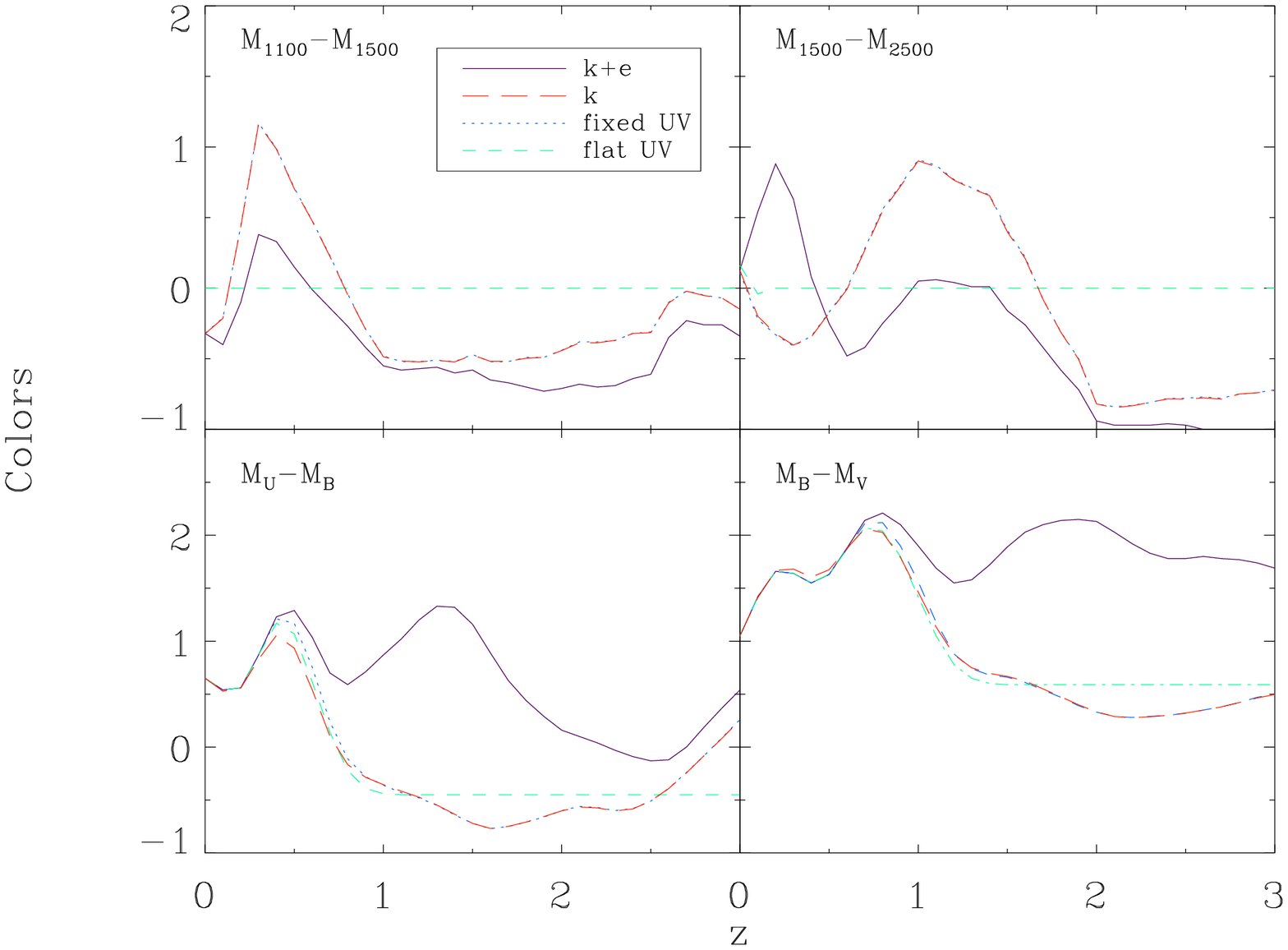}}
\vspace{15pt}
\caption
{
Color evolution. See Figure 1 for the model description. $k$ models and 
$fixed$-$UV$ models are obviously overlapped.
}
\end{figure}

$k$+$e$: This model is both redshift and evolution-corrected. 
Note a sharp rise in the UV brightness (Figure 1) as redshift decreases when 
redshift is small. This happens because a strong UV upturn develops only when 
a galaxy is old enough to contain a substantial number of hot HB stars. 
The rise begins where the dominant UV source changes from PAGB stars to HB 
stars with an increasing age \cite{ydo2}.  
After the rise, the UV brightness increases as we look at a younger and 
brighter (in bolometric luminosity) galaxy.
Such young galaxies have a strong UV flux that mostly comes from PAGB stars, 
but, the relative strength of the UV flux to the optical flux is supposed to 
be much smaller than the present gEs with a UV upturn \cite{ydo1,ydo2}. 
In this model, {\it a UV upturn is never visible in the (observer's frame)
optical band}; the UV upturn in low-$z$ gEs appears still in the observer's
UV while high-$z$ gEs do not have any UV upturn.
(Note: The rate and the initial-point of the UV brightness rise is sensitive 
to input parameters, thus one should not take the particular slope shown in 
Figure 1 too seriously \cite{ydo2}.)

$k$: This model is only redshift-corrected. 
The UV brightness does not decrease with an
increasing redshift much because a strong UV upturn is assumed to exist
always (Figure 1). This model deviates from the $k$+$e$ model significantly at 
$z \gtsim$ 0.7 -- 1 in $UBV$. However, we are not much 
interested in these non-evolving models because stars certainly evolve.

$fixed$-$UV$: A UV upturn is assumed to exist always in this model. 
Its strength, relative to $f(\lambda=2500$\AA), is assumed to remain constant. 
The rest of the spectrum is allowed to evolve normally. This model resembles 
models in which either young MS stars or PAGB models are the dominant
UV source in the sense that the UV upturn phenomenon is not related to age. 
It deviates from the $k$+$e$ model at 
$z \approx$ 0.5 in $U$, at $z \approx$ 0.9 in $B$, and at $z \approx$ 1.2 in 
$V$ (Figure 1). Like $k$ models and $flat$-$UV$ models, $fixed$-$UV$ model 
colors become markedly different from the $k$+$e$ models based on the HB 
hypothesis at about $z \approx$ 1 (Figure 2).

$flat$-$UV$: $f(\lambda<2500$\AA) = $f(\lambda=2500$\AA), i.e. no UV upturn.
This model represents an extreme case where a UV spectrum is flat. Although
not quite the same, UV-weak gEs, e.g., NGC\,3379, show a much flatter slope
in the UV spectrum than NGC\,4552. On a magnitude scale, it is not much 
different from the $fixed$-$UV$ model.

\section*{Conclusions}

   Most of the recent evolutionary population synthesis models have  
suggested that low-mass HB stars are the dominant UV source in gEs: ``the HB 
hypothesis''.
   In such models, a UV upturn is a sign of an old age.
   According to such models, the UV upturn phenomenon would never be
visible in the optical band because, at high-$z$, galaxies are too young to
develop a UV upturn.
   Optical color evolution models based on the HB hypothesis are significantly 
different from simple models, such as $k$-correction-only models or 
$fixed$-$UV$ models, at a moderate redshift ($z \approx$ 0.5 -- 1).
   We should be able to test the models with modern data. 

   It is definitely time to consider an appropriate UV evolution scenario for 
galaxies before a reasonable model for the entire universe is discussed.
   Although several important input parameters are still to be determined,
a simple study clearly demonstrates that neither simple $k$-correction models 
nor $k$+$e$ models with oversimplified UV treatments work when faced with
real data.
   Some studies already seem to support the idea that high-$z$ ($z = 0$ -- 1)
gEs do not show a UV flux that is nearly as strong as that seen in nearby gEs 
\cite{rs}, which is consistent with the HB hypothesis.

\vspace{0.3in}

   This presentation was supported in part by the National Research Council.


\begin{references}

\bibitem{bc} Bruzual, A. G., \& Charlot, S. 1993, ApJ, 405, 538
\bibitem{bur} Burstein, D., Bertola, F., Buson, L. M., Faber, S. M., \& Lauer, T. R. 1988, ApJ, 328, 440
\bibitem{cwb}  Charlot, S., Worthey, G., \& Bressan, A. 1996, ApJ, 457, 625
\bibitem{chi} Chiosi, C., Vallenari, A., \& Bressan, A. 1997, A\&A Suppl., 121, 301
\bibitem{cw} Code, A.D., \& Welch, G.A. 1979, ApJ, 228, 95 
\bibitem{d96} Demarque, P., Chaboyer, B., Guenther, D., Pinsonneault, L., Pinsonneault, M., \& Yi, S. 1996, The Yale Isochrones 1996 (from Yi's web page http://shemesh.gsfc.nasa.gov/astronomy.html)
\bibitem{dor} Dorman, B., O'Connell, R., \& Rood, R. T. 1995, ApJ, 442, 105
\bibitem{gr} Greggio, L., \& Renzini, A. 1990, ApJ, 364, 35
\bibitem{grv} Guiderdoni, B., \& Rocca-Volmerange, B. 1987, A\&A, 186, 1
\bibitem{gst} Gunn, J. E., Stryker, L. L., \& Tinsley, B. M. 1981, ApJ, 249, 48
\bibitem{hdp} Horch, E., Demarque, P., \& Pinsonneault, M. 1992, ApJ, 388, L53
\bibitem{rs} Rakos, K. D., \& Schombert, J. M. 1995, ApJ, 439, 47
\bibitem{tan} Tantalo, R., Chiosi, C., Bressan, A., \& Fagotto, F. 1996, A\&A, 311, 361
\bibitem{ydk} Yi, S., Demarque, P., \& Kim, Y.-C. 1997, ApJ, 482, 677
\bibitem{ydo1} Yi, S., Demarque, P., \& Oemler, A. Jr. 1997, ApJ, 486, in press
\bibitem{ydo2} Yi, S., Demarque, P., \& Oemler, A. Jr. 1997, ApJ, submitted 

\end{references}
\end{document}